\documentclass[prl,longbibliography,amssymb,twocolumn,superscriptaddress]{revtex4-1}
\usepackage{amsmath}
\usepackage{amssymb}
\usepackage{amsthm}
\usepackage{amsfonts}
\usepackage{gensymb} 
\usepackage{listings}
\usepackage{enumerate}
\usepackage{latexsym}
\usepackage{psfrag}
\usepackage{bm}
\usepackage[all]{xy}
\usepackage{graphicx}
\usepackage{subfigure}
\usepackage[pdftex,colorlinks]{hyperref}
\usepackage{color}
\usepackage{mathtools}

\begin{document}

\title{Magnetic Raman continuum in single crystalline H$_3$LiIr$_2$O$_6$}
\author{Shenghai Pei}
\thanks{These two authors contribute to this work equally.}
\affiliation{Department of Physics, Harbin Institute of Technology, Harbin
  150001, China}
\affiliation{Shenzhen Institute for Quantum Science and Engineering, and Department of Physics, Southern University of Science and Technology, Shenzhen 518055, China}
\author{Liang-Long Huang}
\thanks{These two authors contribute to this work equally.}
\affiliation{Shenzhen Institute for Quantum Science and Engineering, and Department of Physics, Southern University of Science and Technology, Shenzhen 518055, China}

\author{Gaomin Li}
\affiliation{School of Advanced Materials, Shenzhen Graduate School Peking
  University, Shenzhen 518055, P. R. China}
\affiliation{Shenzhen Institute for Quantum Science and Engineering, and
  Department of Physics, Southern University of Science and Technology, Shenzhen
  518055, China}
\author{Xiaobin Chen}
\affiliation{School of Science, Harbin Institute of Technology, Shenzhen 518055, China}

\author{Bin Xi}
\affiliation{College of Physics Science and Technology, Yangzhou University, Yangzhou 225002, China}
\author{XinWei Wang}
\affiliation{School of Advanced Materials, Shenzhen Graduate School Peking University, Shenzhen 518055, P. R. China}

\author{Youguo Shi}
\affiliation{Institute of Physics, Chinese Academy of Sciences, Beijing 100190,
  China}
\affiliation{School of Physical Sciences, University of Chinese Academy of Sciences, Beijing 100190, China}

\author{Dapeng Yu}


\author{Cai Liu}
\author{Le Wang}
\email{wangl2018@mail.sustc.edu.cn}
\author{Fei Ye}

\affiliation{Shenzhen Institute for Quantum Science and Engineering, and
  Department of Physics, Southern University of Science and Technology, Shenzhen
  518055, China}

\author{Mingyuan Huang}
\email{huangmy@sustc.edu.cn}
\affiliation{Shenzhen Institute for Quantum Science and Engineering, and Department of Physics, Southern University of Science and Technology, Shenzhen 518055, China}
\affiliation{Shenzhen Key Laboratory of Quantum Science and Engineering,
  Shenzhen 518055, PR China.}
  
\author{Jia-Wei Mei}
\email{meijw@sustc.edu.cn}
\affiliation{Shenzhen Institute for Quantum Science and Engineering, and Department of Physics, Southern University of Science and Technology, Shenzhen 518055, China}

\date{\today}

\begin{abstract}
Recently H$_3$LiIr$_2$O$_6$ has been reported as a spin-orbital entangled quantum
spin liquid (QSL) [K. Kitagawa et al., Nature {\bf 554}, 341 (2018)], albeit its
connection to Kitaev QSL has not been yet identified. To unveil the related
Kitaev physics, we perform the first Raman spectroscopy studies on single
crystalline H$_3$LiIr$_2$O$_6$ samples. We implement  a soft chemical
replacement of Li$^+$ with H$^+$ from $\alpha$-Li$_2$IrO$_3$ single
crystals to synthesize the single crystal samples of the iridate second generation H$_3$LiIr$_2$O$_6$. The Raman spectroscopy can be used to diagnose the
QSL state since the magnetic Raman continuum arises from a process
involving pairs of fractionalized Majorana fermionic excitation in a pure Kitaev
model. We observe a broad dome-shaped magnetic continuum in H$_3$LiIr$_2$O$_6$, in line with
theoretical expectations for the two-spin process  in the Kitaev QSL. Our
results establish the close connection to the Kitaev QSL physics in H$_3$LiIr$_2$O$_6$.
\end{abstract}

\maketitle

\emph{Introduction. --}
The search for quantum spin liquid (QSL) state has been a currently active and challenging topic in the condensed
matter
physics~\cite{Anderson1973,Anderson1987,Balents2010,Savary2017,Zhou2017,Shimizu2003,Nytko2008,Feng2017,Wen2017,Wei2017,Feng2018,Kitagawa2018}.
The spin degree of freedom in QSL does not freeze to display any magnetic order even at
zero temperature, but highly entangles with each
other~\cite{Kitaev2006,Wen2004,Kitaev2006a,Levin2006,Wen2017a,Wen2019}.
In the early theoretical studies, the
exact solvable Kitaev honeycomb spin model~\cite{Kitaev2006}  built  confidence
about the existence of QSL in a simple spin interacting system; furthermore, it has been currently initialing the materialization of the
Kitaev QSL in the experiments~\cite{Jackeli2009,Chaloupka2010,Takagi2019}. With the help of the intertwining between magnetism,
spin-orbital coupling, and crystal field,  Ir$^{4+}$ oxides and a Ru$^{3+}$ chloride with
a $d^5$ electronic configuration are promising to materialize the Kitaev
model, \textit{e.g.}, $\alpha$-A$_2$IrO$_3$ (A=Na,
Li)~\cite{Choi2012,Singh2012,HwanChun2015} and
$\alpha$-RuCl$_3$~\cite{Plumb2014}.

Due to other non-Kitaev
interactions, magnetic orders appears in $\alpha$-A$_2$IrO$_3$ (A=Na,
Li) and $\alpha$-RuCl$_3$ at low
temperatures~\cite{Choi2012,Johnson2015,Williams2016}. 
The suppression of magnetic ordering in $\alpha$-A$_2$IrO$_3$ (A=Na,
Li) and $\alpha$-RuCl$_3$ has been attempted by applying magnetic
field~\cite{Kasahara2018a,Yu2018}
, high pressure~\cite{Hermann2018,Li2019,Note2}, and
chemical modification~\cite{Bette2017,Kitagawa2018}. 
For the chemical modification of $\alpha$-Li$_2$IrO$_3$, a QSL state ground was recently established in the
second generation of two-dimensional honeycomb iridates
H$_3$LiIr$_2$O$_6$~\cite{Bette2017,Kitagawa2018}. No
sign of magnetic order,  but signatures
of local low-energy excitations are observed in H$_3$LiIr$_2$O$_6$, down to low temperatures in the magnetic susceptibility,
specific heat, and NMR measurements~\cite{Kitagawa2018}. H$_3$LiIr$_2$O$_6$ has immediately
caught lots of theoretical investigation to explore the connection to the Kitaev
QSL physics~\cite{Slagle2018,Yadav2018,Li2018,Wang2018,Knolle2019}, and the ranomness of H positions was also discussed as
playing an important role in stabilizing the QSL state~\cite{Yadav2018,Li2018}.
Currently, however, no spectroscopic information exists regarding the spin
excitations and the possibility of spin fractionalization in H$_3$LiIr$_2$O$_6$. 

In this Letter, we report our attempts to diagnose the spin liquid signature in
the single crystalline H$_3$LiIr$_2$O$_6$ using the Raman spectroscopy methods.
Single crystals of $\alpha$-Li$_2$IrO$_3$ are soaked in 4 mol/L H$_2$SO$_4$
aqueous solution for the soft chemical replacement of Li$^+$ with H$^+$. As well
as single crystals of the target second generation of iridate
H$_3$LiIr$_2$O$_6$, we obtain the third iridate
generation with the hypothesized formula H$_5$LiIr$_2$O$_6$. We carry out
the X-ray photoelectron spectroscopy (XPS) measurements, and confirm that
H$_3$LiIr$_2$O$_6$ has the same Ir$^{4+}$ oxidation state as
$\alpha$-Li$_2$IrO$_3$, while H$_5$LiIr$_2$O$_6$ has lower oxidation state
Ir$^{3+}$.  Submillimetre-size crystals are available for the Raman
spectroscopy, that is capable of detecting magnetic excitations
~\cite{Moriya1968,Fleury1968,Lemmens2003}, even the spin fractionalizations
signaled by the magnetic Raman continuum in the Kitaev-type
compounds\cite{Sandilands2015,Glamazda2016,Glamazda2017,Li2019,Note2}. We 
observe a broad two-spin process continuum in the dynamical Raman susceptibility
for H$_3$LiIr$_2$O$_6$, in a good
agreement with the theoretically expected
scattering from a pure Kitaev model~\cite{Knolle2014}.  Our results demonstrate Raman spectroscopic
signautres of the fractionlized excitation for the Kiteav QSL state in H$_3$LiIr$_2$O$_6$.

\textit{Single crystal synthesis and experimental setup. --}
We implement the soft chemical replacement of of Li$^+$ with H$^+$ in the
iridate first generation  $\alpha$-Li$_2$IrO$_3$ single crystals. We
obtain single crystals  of different generations of honeycomb iridate
orxides, depending the
growth condition, especially reaction (soaking) time. For the
growth of $\alpha$-Li$_2$IrO$_3$ single crystal, we adopted the similar setup as described in
Ref.~\cite{Freund2016} with iridium metal (powder, 99.99\%) and lithium
(granule, 99.99\%) used as starting materials. The whole setup was placed in a
preheated furnace of 200~\celsius~ heated to 1020~\celsius~ and dwelled for
10~days. For cation exchange,
$\alpha$-Li$_2$IrO$_3$ single crystals were added into in a 25~ml Teflon-lined
steel autoclave with 20~ml H$_2$SO$_4$ aqueous solution (4~mol/L).

The XPS investigations were carried out on Thermo Fisher ESCALAB 250Xi using
monochromated Al K$\alpha$ radiation at room temperature, and the electron flood gun was
turned on to eliminate electric charging effect in our insulating samples. The
binding energy in XPS was calibrated by 1s spectra of carbon. 
XRD measurements were conducted on Rigaku Smartlab 9KW using Cu K$\alpha$
radiation at room temperature. 

\begin{figure}[b]
  \centering
  \includegraphics[width=\columnwidth]{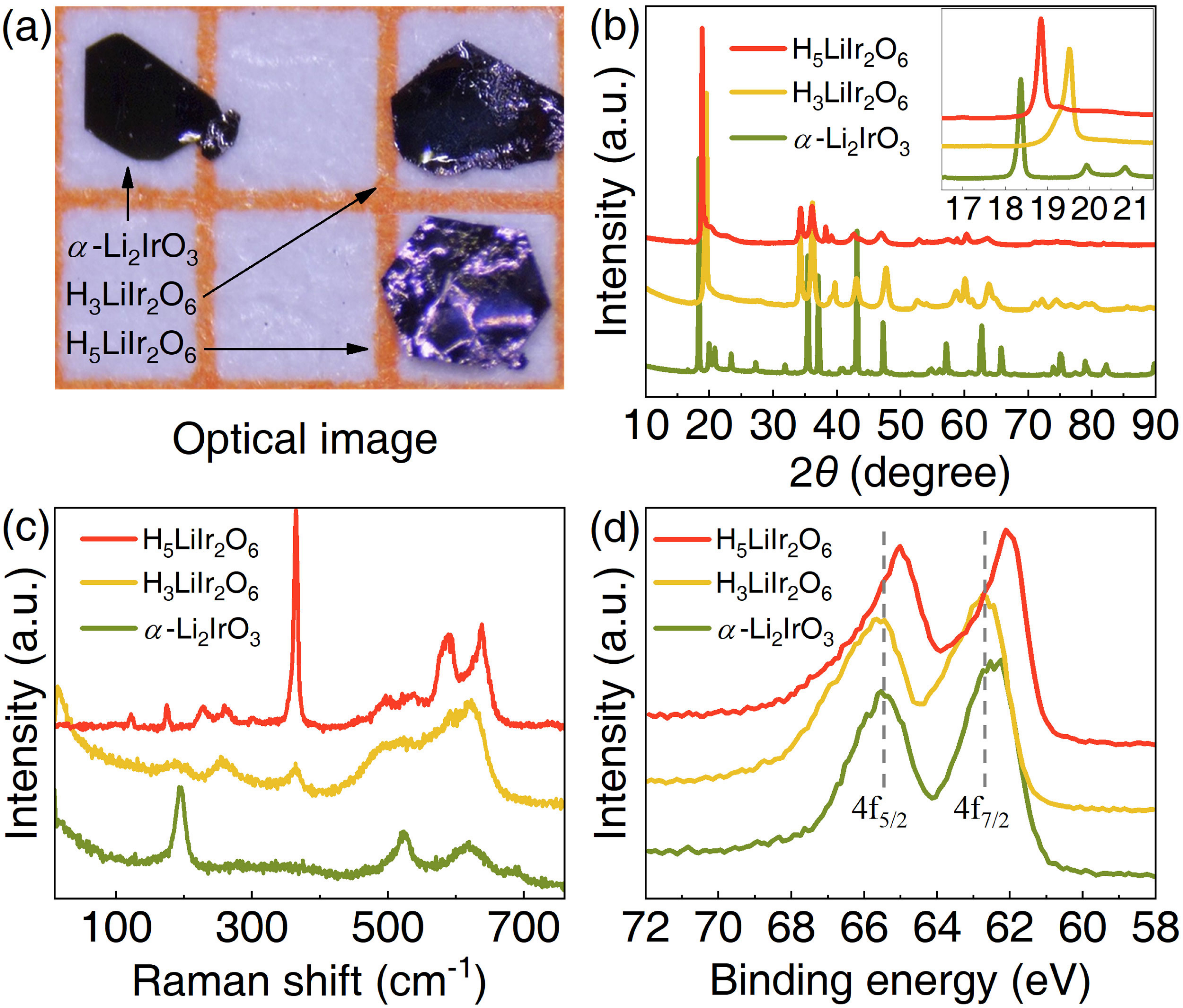}
  \caption{(a) Image of single crystals for $\alpha$-Li$_2$IrO$_3$,
    H$_3$LiIr$_2$O$_6$ and H$_5$LiIr$_2$O$_6$, respectively. The yellow
    background grid is 1$\times$1~mm$^2$. To characterize
    three iridate generations at a room temperature, we have measured PXRD
    patterns in (b), Raman spectra in (c), and XPS in (d).}
  \label{fig:figure1}
\end{figure}
The Raman spectra were measured in the quasi-back-scattering geometry, with
light polarized in the basal plane. The experiments were performed on our home-built system 
using a HORIBA iHR550 spectrometer and the 632.8~nm excitation line of a He-Ne laser. The power of the laser
was kept low enough (about 100~$\mu$W) to prevent from heating and damaging
samples. We use 1200~grooves/mm grating to get the high resolution. Since
the light scattering intensity is weak, we set the integral time to 1800~s. The samples were
placed in a He-flow cryostat which evacuated to $2.0\times10^{-6}$~Torr. The sample temperatures were calibrated according to the
intensity ratio of anti-Stokes and Stokes phonon peaks.

\emph{Sample characterizations. --} 
Single crystals of $\alpha$-Li$_2$IrO$_3$ have different appearances, e.g., the
flake and pyramid shapes, probably due to different stacking patterns of
LiIr$_2$O$_6$ layers. We find that the pyramid-shaped $\alpha$-Li$_2$IrO$_3$
have more ordered stacking pattern, however, the flake-shaped crystals are easily
accessible for the soft chemical replacement as the present study in this work. 
Fig.~\ref{fig:figure1}~(a) is the image for typical single crystals of three generations of
the iridate oxides. The parent generation $\alpha$-Li$_2$IrO$_3$ (the flake cyrstal) displays the
black appearance. The second generation (H$_3$Li$_2$O$_6$)
has the reddish black color, and the third generation (hypothesized formula H$_5$LiIr$_2$O$_6$) has the
lustrous red appearance. During the soft-chemical-ion-exchange reaction (about 70~min), the interlayer Li$^+$ will be
replaced by H$^+$, and we
can get the target iridate second generation H$_3$LiIr$_2$O$_6$. With longer soaking time (about 3~hours), more Hydrogen
atoms intercalate into the inter-layers of [LiIr$_2$O$_6$] layers, and we get
the crystals of the third generation H$_5$LiIr$_2$O$_6$. 
While  $\alpha$-Li$_2$IrO$_3$ and H$_5$LiIr$_2$O$_6$ is
very stable, H$_3$LiIr$_2$O$_6$ may react with the vapour in the air.

Figure.~\ref{fig:figure1}~(b) is the powder X-ray diffraction (PXRD) result for three generations of iridate
oxides. Apparently, they have very similar overall PXRD patterns since  the acid treatment have caused a
mild change in the constitution of the LiIr$_2$O$_6$ layers, adn only the ions
situated in-between the layers are changed. 
The intense basal reflection $2\theta$  shifts up from $\alpha$-Li$_2$IrO$_3$
(18.36\degree) to H$_3$LiIr$_2$O$_6$ (19.52\degree), and then down to
H$_5$LiIr$_2$O$_6$ (18.85\degree), corresponding to the interlayer distance 4.828~\AA,
4.544~\AA~ and 4.704~\AA, respectively. Strong anisotropic broadening of the reflections in
the XRD-pattern revealed heavy stacking faulting of the samples.
Fig.~\ref{fig:figure1}~(c) is the typical Raman result for the three generations
at a room temperature. The Raman intensity of of H$_5$LiIr$_2$O$_6$ rescales by
multiplying by 1/2. During the crystal synthesis, we use the Raman spectra to monitor the soaking process of
our samples.

We implement XPS to verify the oxidation state of iridium ions in the three
generations.  As shown in Fig.~\ref{fig:figure1}~(d), three
generations have very similar Ir 4$f$ XPS spectra, indicating the similar local
electronic environment of IrO$_6$. The shift of the binding energy of Ir 4$f$
implies the changes of oxidation state of Ir,  and lower binding
energy represents lower valance state. The mild change can be seen from
$\alpha$-Li$_2$IrO$_3$ to H$_3$LiIr$_2$O$_6$ by comparing the position of
4f$_{5/2}$ and 4f$_{7/2}$ energy level, indicating the same electronic
configuration of Ir$^{4+}$. H$_5$LiIr$_2$O$_6$ has a significant lower binding
energy (about 0.7~eV). According to Ir-based compounds in the
database\footnote{https://srdata.nist.gov/xps/Default.aspx}, Ir$^{3+}$ and
Ir$^{4+}$ usually have a difference of the binding energy about 0.5~eV.
Comparing Ir$^{4+}$ in $\alpha$-Li$_2$IrO$_3$ and H$_3$LiIr$_2$O$_6$, we have
Ir$^{3+}$ in IrO$_6$ octahedron in H$_5$LiIr$_2$O$_6$. Ir$^{3+}$ has the 3$d^6$
electronic configurations without partially occupied orbitals, and hence no
magnetism. This explains the negligible magnetic Raman continuum in
H$_5$LiIr$_2$O$_6$ as shown in Fig.~\ref{fig:figure1}~(c) and Fig.~\ref{fig:figure3}.

\begin{figure}[t]
\centering
\includegraphics[width=\columnwidth]{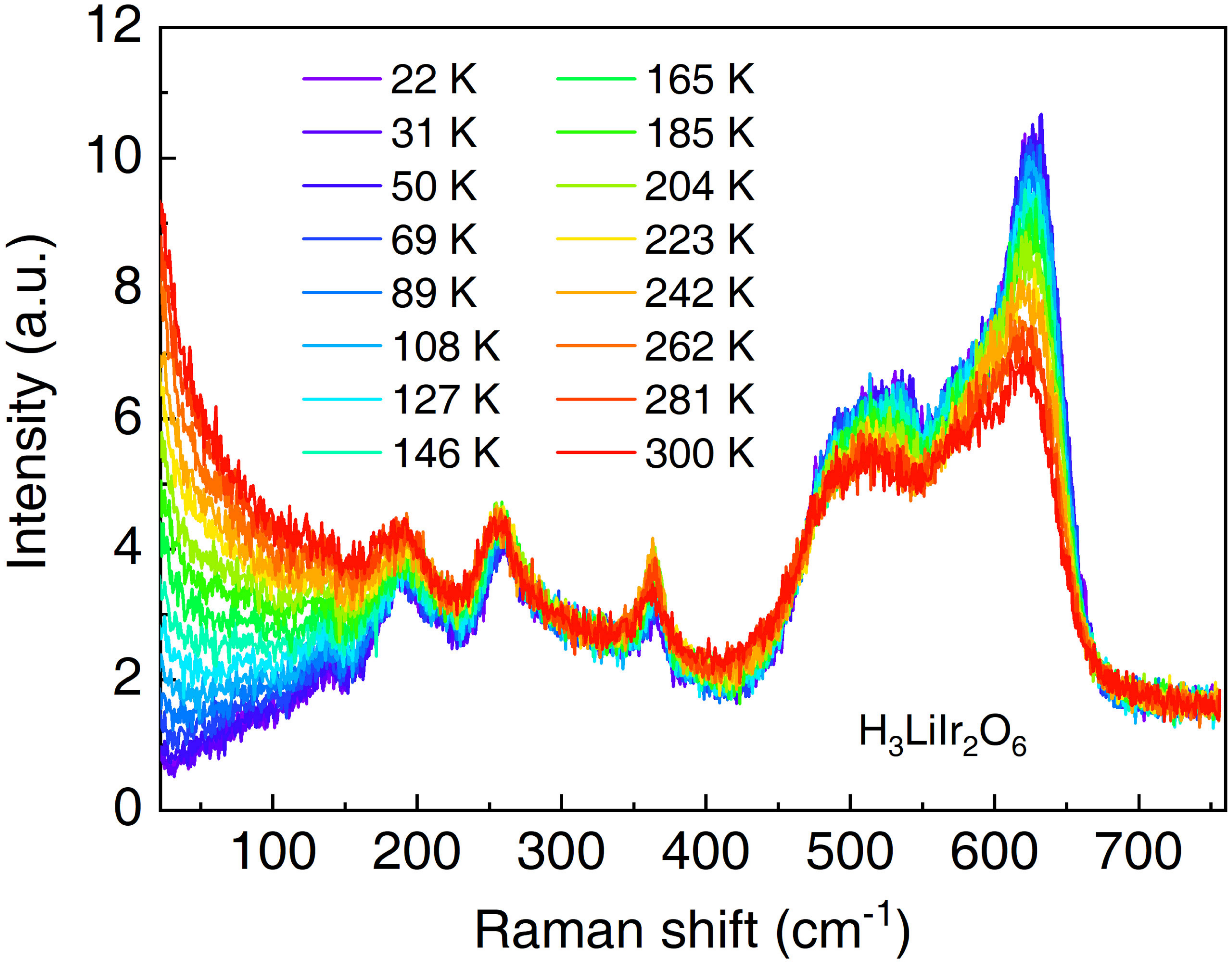}
\caption{Evolution of the Raman spectra at different temperatures in
  H$_3$LiIr$_2$O$_6$. The spectra composes sharp phonon peaks and the broad magnetic
continuum.}
\label{fig:figure2}
\end{figure}

\emph{Evolution of Raman spectra --}
Figures.~\ref{fig:figure2}  is the evolution of Raman
spectra of H$_3$LiIr$_2$O$_6$ at different temperatures.
Comparing with our
$\alpha$-Li$_2$IrO$_3$ results~\footnote{Gaomin Li, \textit{et. al.}, to be
  submitted.}, we can assign the sharp phonon modes by assuming
that H$_3$LiIr$_2$O$_6$ has the same space group (\#12, C2/m) as $\alpha$-Li$_2$IrO$_3$.
Three modes at 364.0 and 630.8~cm$^{-1}$ are assigned as $A_g$ modes, the
mode at 131.9~cm$^{-1}$ is assigned as $B_g$ mode, and four modes at 189.6,
259.9, 490.0 and 541.0~cm$^{-1}$ are assigned as $A_g+B_g$ modes which are nearly
doubly degenerate.  131.9 and 189.6~cm$^{-1}$ modes are the Ir-Ir out-of-phase
motions along the out-of-plane and in-plane directions, respectively. The
259.9~cm$^{-1}$ mode  is the twist of Ir-O-Ir-O plane, the 364.0 cm-1 mode is
the relative twist of between the upper and lower oxygen triangles. The 490~cm$^{-1}$ mode is  related
to the Ir-O-Ir-O plane shearing, and the 541.0~cm$^{-1}$ mode is the breathing
mode of Ir-O-Ir-O ring. The Ag mode at 630.8~cm$^{-1}$ can be assigned as the
symmetrical breathing mode between the upper and lower oxygen layers. Several
weak phonon peaks at around 155, 220, 325 and 600~cm$^{-1}$ become visible at low temperatures in
H$_3$LiIr$_2$O$_6$, and these modes don't appear in
$\alpha$-Li$_2$IrO$_3$~\cite{Note2}. We notice
that Raman phonon modes in H$_3$LiIr$_2$O$_6$ in this work
and $\alpha$-Li$_2$IrO$_3$~\cite{Note2} are quite similar to those in other
Kitaev materials, e.g. $\alpha$-RuCl$_3$~\cite{Sandilands2015,Li2019} and $\beta$- and $\gamma$-
Li$_2$IrO$_3$~\cite{Glamazda2016}, suggestive of similar local crystal RuCl$_6$ and
IrO$_6$ octahedral structures.

\begin{figure}[t]
  \centering
  \includegraphics[width=\columnwidth]{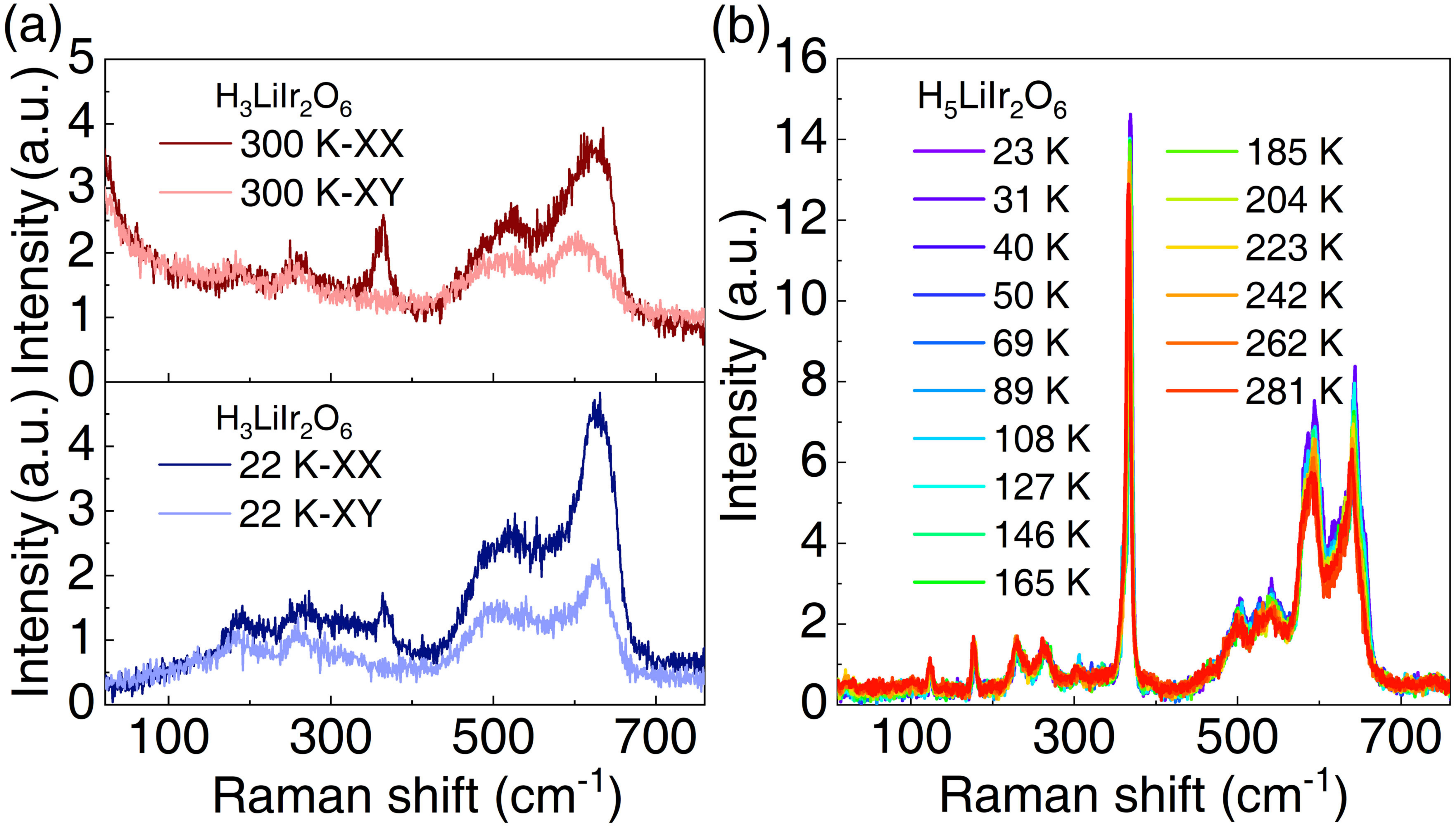}
  \caption{(a) The polarized Raman spectra at 300~K (upper) and 20~K in
    H$_3$LiIr$_3$O$_6$. (b) Evolution of the Raman spectra at different
    temperatures in H$_5$LiIr$_2$O$_6$. As a control experiment, there is no magnetic Raman continuum
    at all. }
  \label{fig:figure3}
\end{figure}
Besides the phonon modes, we observe a strong continuum background with
increasing intensity as increasing temperatures particularly for low Raman frequencies (Fig.~\ref{fig:figure2}). We attribute such a continuum background as the magnetic Raman scattering as observed
in  $\alpha$-RuCl$_3$~\cite{Sandilands2015} and $\beta$- and
$\gamma$-Li$_2$IrO$_3$~\cite{Glamazda2016} due to spin fractionalized
excitations. H$_3$LiIr$_2$O$_6$ is a layered material and there are heavily stacking faults which may lead to
broad phonon modes that could explain the continuum. 
However, the static structural disorder cannot
produce the significant temperature dependence. Furthermore, the damped phonon scenario
couldn't account for the Raman continuum according to the polarization
dependence of the Raman spectra in H$_3$LiIr$_2$O$_6$ as shown in
Fig.~\ref{fig:figure3}~(a). As a matter of fact, the weak polarization
dependence of Raman spectra in H$_3$LiIr$_2$O$_6$ agrees well with theoretical
calculations for the isotropic Kitaev model~\cite{Knolle2014a}.

As a control experiments, Fig.~\ref{fig:figure3}~(b) is the evolution of Raman
spectra in H$_5$LiIr$_2$O$_6$ at different
temperatures. Actually, H$_5$LiIr$_2$O$_6$ would have similar stacking faults to
H$_3$LiIr$_2$O$_6$ since they main difference in their synthesis is the soaking
time of the soft chemical reaction.  With increasing temperature, all phonon
peaks in H$_5$LiIr$_2$O$_6$  change very mild, and
there is not any continuum background at all at the whole temperature range, which is in contrast to the
spectra of H$_3$LiIr$_2$O$_6$ in Fig.~\ref{fig:figure2}. On one hand, as a
control experiment, the Raman spectra of H$_5$LiIr$_2$O$_6$ in
Fig.~\ref{fig:figure3}~(b) implies that the continuum background of H$_3$LiIr$_2$O$_6$ in
Fig.~\ref{fig:figure2} is not due to the stacking faults, but has the magnetic
origin. On the other hand, even if some regions of H$_3$LiIr$_2$O$_6$
(Ir$^{4+}$) turn into H$_5$LiIr$_2$O$_6$ (Ir$^{3+}$) due to the over soaking in the
acid, they behave as handful impurities and do not contribute to the magnetic Raman continuum.

\begin{figure}[t]
\centering
\includegraphics[width=\columnwidth]{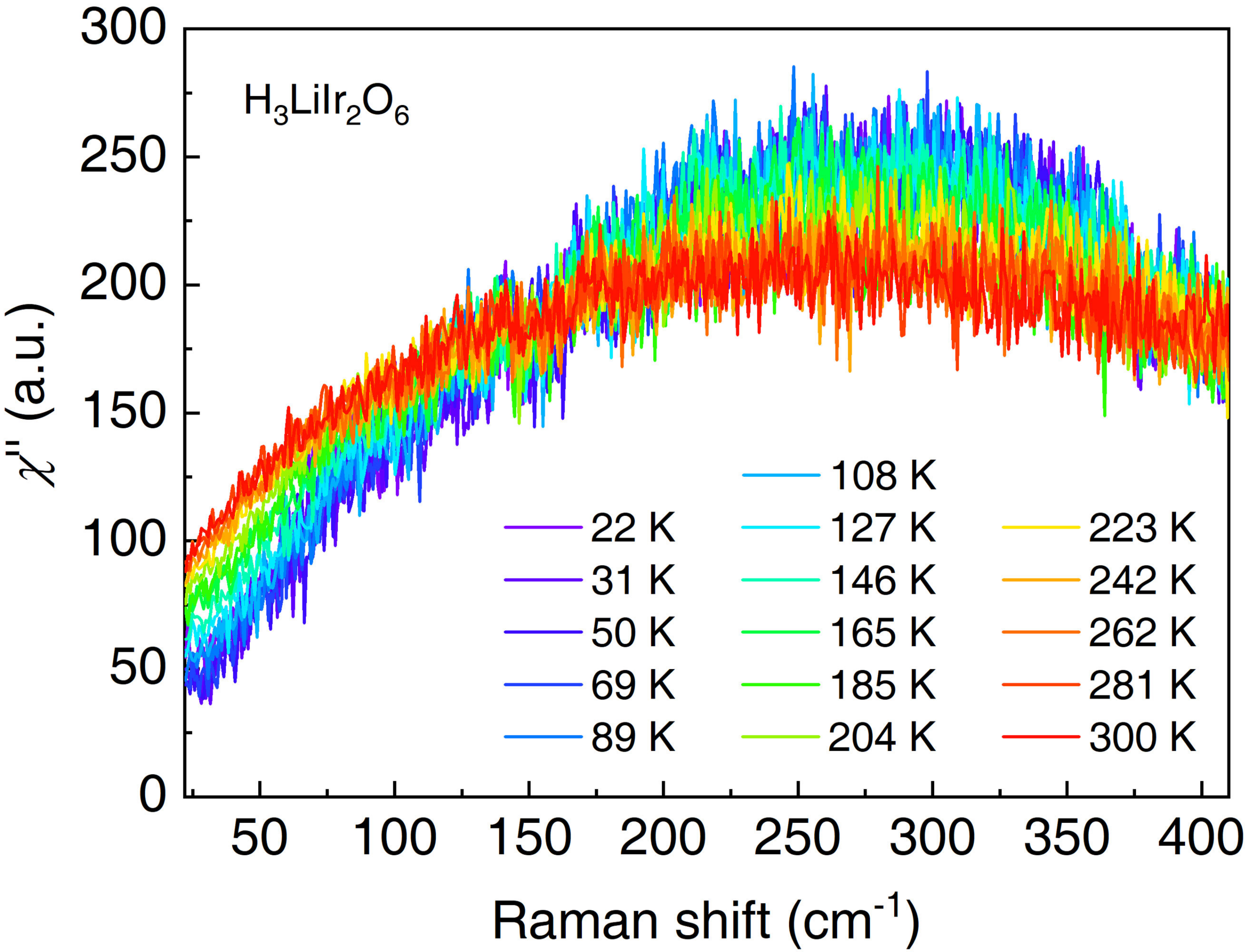}
\caption{Temperature dependent revolution of magnetic Raman continuum in the
  dynamical Raman susceptibility $\chi''(\omega)$ in
  H$_3$LiIr$_2$O$_6$. }
\label{fig:figure4}
\end{figure}
\begin{figure}[b]
  \centering
  \includegraphics[width=0.8\columnwidth]{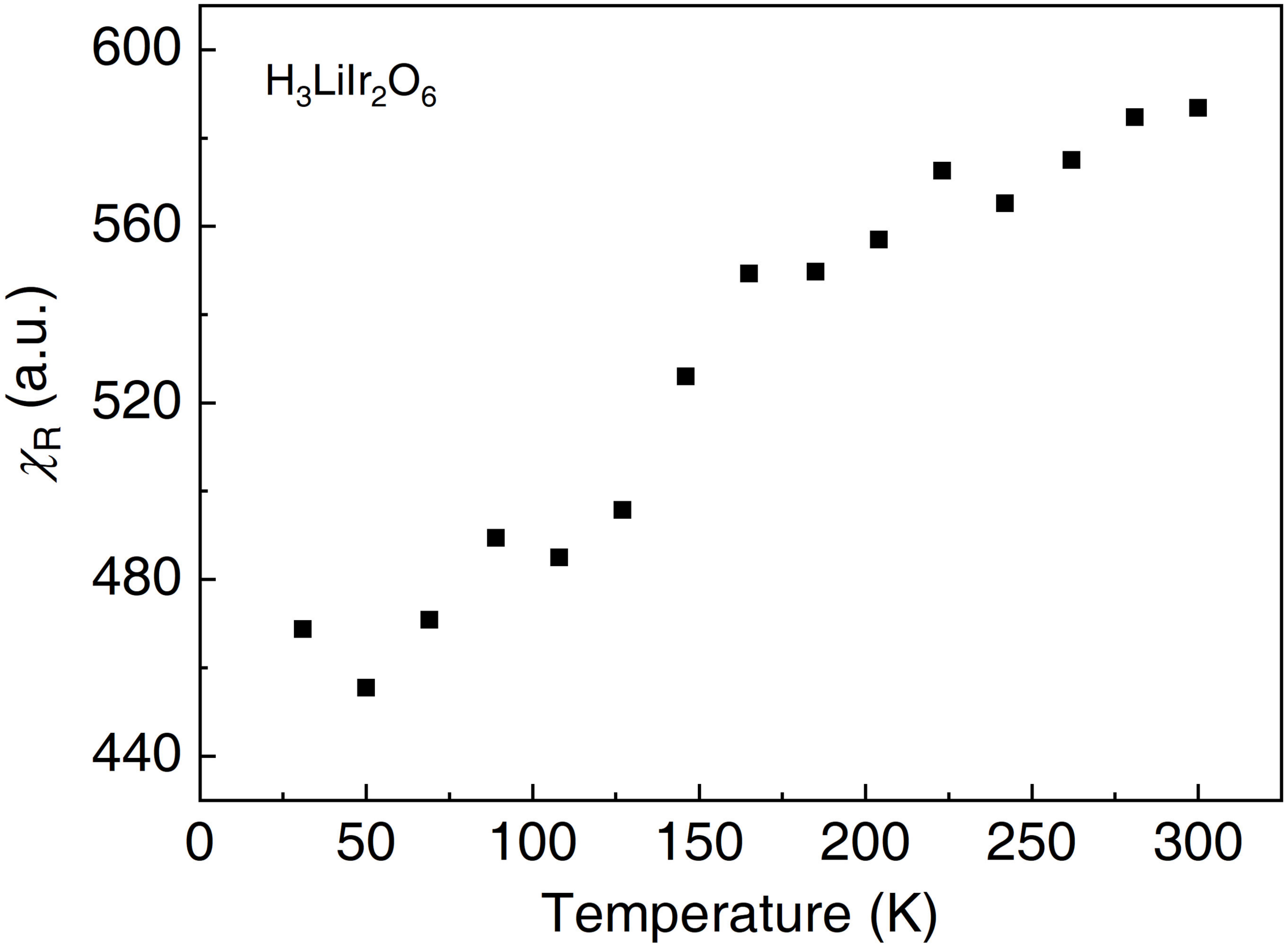}
  \caption{Temperature dependent magnetic Raman susceptibility $\chi_R(T)$ in
    H$_3$LiIr$_2$O$_6$.}
  \label{fig:figure5}
\end{figure}
According to the fluctuation-dissipation theorem, the
Raman intensity $I(\omega)$ is proportional to the dynamical Raman
 susceptibility as
$I(\omega) = [1+n(\omega)]\chi''(\omega)$. Here $n(\omega)$ is the boson factor,
and $\chi''(\omega)$
is the imaginary part of the correlation functions of Raman tensor
$\tau(\mathbf{r},t)$,
$\chi(\omega)=\int_{0}^{\infty}dt\int
d\mathbf{r}(-i)\langle[\tau(0,0),\tau(\mathbf{r},t)] \rangle e^{-i\omega
  t}$. To examine the  the magnetic Raman susceptibility $\chi''(\omega)$ in
H$_3$LiIr$_2$O$_6$ more
explicitly, we remove phonon modes using the Gaussian-type line-shape. As a
consequence, the obtained magnetic Raman susceptibility in
H$_3$LiIr$_2$O$_6$ displays a dome-shaped
broad continuum at all temperatures, as shown in Fig.~\ref{fig:figure4}. With the temperature decreasing, the magnetic Raman continuum
increases with the frequency between 150~cm$^{-1}$ and 410~cm$^{-1}$, and
decreases with the frequency less than 150~cm$^{-1}$. Thus the dome shape of the
magnetic continuum is more remarkable at low temperatures. From the Raman
intensity in Fig.~\ref{fig:figure2}, we can see that the magnetic continuum
extends to higher frequencies at least up to 700~cm$^{-1}$. The phonon
modes with the frequency between 410~cm$^{-1}$ and 710~cm$^{-1}$ are messy, and
it is not easy to separate the phonon modes and magnetic continuum in the frequency region
(between 410~cm$^{-1}$ and 710~cm$^{-1}$). Therefore, we didn't show the
subtracted magnetic Raman continuum with the frequency between 410~cm$^{-1}$ and
710~cm$^{-1}$ in Fig.~\ref{fig:figure4}. We can see that the magnetic continuum is weakly temperature dependent in this
frequency region according to the Raman spectra in H$_3$LiIr$_2$O$_6$ in Fig.~\ref{fig:figure2}. 

We extract the integrated Raman susceptibility $\chi_R$ in accordance with the Kramers-Kronig
relation $\chi_R=\frac{2}{\pi}\int\frac{\chi''(\omega)}{\omega}d\omega$. To do
the integration, we extrapolate the Raman conductivity
$\frac{\chi''(\omega)}{\omega}$ in
H$_3$LiIr$_2$O$_6$ to 0~cm$^{-1}$. The temperature dependent
$\chi_R$ in
H$_3$LiIr$_2$O$_6$ is plotted in Fig.~\ref{fig:figure5}, in which the integration is
carried out from 0~cm$^{-1}$ to 410~cm$^{-1}$, taking into account the fact
that the magnetic Raman continuum with the frequency between 410~cm$^{-1}$ and
710~cm$^{-1}$ is weakly temperature dependent. The integrated Raman
susceptibility $\chi_R$ in H$_3$LiIr$_2$O$_6$ essentially decreases
monotonically as lowering the temperatures at least above 50~K, very different
from that in $\alpha$-Li$_2$IrO$_3$\cite{Note2}, where $\chi_R$ increases
monotonically with temperature decreasing.  It is worthy to
mention that the temperature dependence of $\chi_R$ in H$_3$LiIr$_2$O$_6$ is similar
to $\beta$- and $\gamma$-Li$_2$IrO$_3$~\cite{Glamazda2016}.  The
integrated Raman susceptibility $\chi_R$ of $\alpha$-Li$_2$IrO$_3$ has a very
similar temperature dependence behavior  to $\alpha$-RuCl$_3$, \textit{i.e.} increasing
monotonically with temperature decreasing~\cite{Note2,Glamazda2017}.
More
specifically, the inelastic light scattering in H$_3$LiIr$_2$O$_6$ and $\beta$-
and $\gamma$-Li$_2$IrO$_3$ has different form from that in $\alpha$-RuCl$_3$ and
$\alpha$-Li$_2$IrO$_3$, which deserves further detail investigations.

\emph{Discussions and conclusions --}
With the help of strong spin-orbit coupling, crystal field splitting and
electronic correlation, the Kitaev materials have the effective spin-1/2
moment~\cite{Jackeli2009,Takagi2019}. The magnetic Raman tensor
$\tau(\mathbf{r})$ in these systems can be expanded in powers of the effective spin-1/2 operators,
$\tau^{\alpha\beta}(\mathbf{r})=\tau_0^{\alpha\beta}(\mathbf{r})+\sum_\mu
K^{\alpha\beta}_{\mu}S^\mu(\mathbf{r})+\sum_{\delta}\sum_{\mu\nu}M^{\alpha\beta}_{\mu\nu}(\mathbf{r},\delta)S_{\mathbf{r}}^\mu
S_{\mathbf{r}+\delta}^\nu+...$. The first term corresponds to Rayleigh
scattering, the second and third terms correspond to the one-spin and
two-spin process, respectively~\cite{Moriya1968,Fleury1968,Lemmens2003}.
The complex tensors $K^{\alpha\beta}_{\mu}$ and $M^{\alpha\beta}_{\mu\nu}$ are determined by the strength of the spin-orbit
couplings and the subtle coupling form of light to the spin system. If the
one-spin process dominates the inelastic light scattering, the integrated
Raman susceptibility $\chi_R$ is associated with the thermodynamic magnetic
susceptibility $\chi$, as demonstrated in $\alpha$-RuCl$_3$~\cite{Glamazda2017}
and $\alpha$-Li$_2$IrO$_3$~\cite{Note2}. In $\beta$- and $\gamma$-Li$_2$IrO$_3$,
the integrated Raman susceptibility
$\chi_R^0$ is associated with the magnetic-specific heat $C_m$ multiplied by the temperature
$T$, \textit{i.e.,} $C_mT$~\cite{Glamazda2016}, indicating that the two-spin
process dominates the magnetic Raman scattering. 

The temperature dependence behavior of  H$_3$LiIr$_2$O$_6$ is similar to that in $\beta$-
and $\gamma$-Li$_2$IrO$_3$, therefore, two-spin process dominates in the magnetic Raman continuum
in Fig.~\ref{fig:figure4}. In the putative Kitaev QSL, the magnetic Raman
scattering of two-spin process directly probes the pairs of the Majorana
fermions which are characterization of the elusive spin
fractionalizations~\cite{Knolle2014a,Nasu2016}. Particularly, Knolle~\textit{et.
al.} have calculated the magnetic Raman scattering for the two-spin process, and
our dome-shaped magnetic Raman continuum agrees very well with the simulated
Raman response~\cite{Knolle2014a}. Therefore, our results demonstrate the
emergence of spin fractionalization, and establish a Kitaev QSL in H$_3$LiIr$_2$O$_6$. The theoretical
simulated Raman response has a maximum at 1.5~$J_k$ (where $J_k$ is the Kitaev
interaction for the effective spin-1/2 operators)~\cite{Knolle2014a}. Equating 1.5~$J_K$
 with the experimental maximum of continuum scattering of 40~meV yields
 $J_K=26$~meV, in a good consistent with DFT
 estimations~\cite{Yadav2018,Li2018,Wang2018}.

 H$_3$LiIr$_2$O$_6$ have heavily stacking faults~\cite{Kitagawa2018,Bette2017}, and theoretical investigations of  have discussed the important
 role of the randomness to stabilize the Kitaev
 QSL~\cite{Slagle2018,Yadav2018,Li2018,Wang2018,Knolle2019}. However, Raman
 scattering is not sensitive to the local randomness, and our results didn't
 address the issue about the physics of disorder in quantum
 spin liquids. Knolle~\textit{et. al.} predicted a $\delta$-function peak
 reflecting the local two-particle density of states of Majorana fermions in the presence of four
 vison fluxes~\cite{Knolle2014a}. Such a vison peak is not resolved in our Raman data due to the
 severely bad  resolution at small wave numbers.

 In conclusion, we perform the Raman spectroscopy studies of single-crystal
 H$_3$LiIr$_2$O$_6$ samples and observe a dome-shaped magnetic Raman continuum.
 Our results demonstrate the spin fractinalization in H$_3$LiIr$_2$O$_6$, which is
 a defining feature of the Kitaev quantum spin liquid state.

\acknowledgements{\textit{Acknowledgments --} 
  J.W. M thanks L. Zhang for informative discussions, and Y.L. He and A. Ng for
  useful discussions on the XPS measurements. 
  This work was supported by the Science, Technology and
  Innovation Commission of Shenzhen Municipality (Grant
  No.ZDSYS20170303165926217). M.H. was partially supported by the
  Science, Technology and Innovation Commission of Shenzhen Municipality
  (Grant No. JCYJ20170412152334605).  F.Y. was
  partially supported by National Nature Science Foundation of China
  (Grant No. 11774143) and the
  Science, Technology and Innovation Commission of Shenzhen
  Municipality (Grant No. JCYJ20160531190535310). J.W.M was partially supported by the program for Guangdong Introducing Innovative and Entrepreneurial Teams (No. 2017ZT07C062) }

\bibliography{../LiIrO}
\end{document}